\documentclass[aps,prl,twocolumn,superscriptaddress,longbibliography]{revtex4-2}
\usepackage{graphicx}
\usepackage{amssymb,amsmath}
\usepackage{bm}
\usepackage{dcolumn}
\usepackage{float}
\usepackage[OT1]{fontenc} 
\usepackage{url}
\usepackage{mathrsfs}
\usepackage{slashed,comment}
\usepackage{color}
\usepackage{verbatim}
\usepackage{enumitem}
\usepackage{soul,physics}
\usepackage[driverfallback=dvipdfm]{hyperref}
\hypersetup{pdfpagemode=FullScreen,colorlinks=true,breaklinks,urlcolor=blue,linkcolor=blue,citecolor=blue}

\usepackage{subfigure}
\usepackage{amssymb}
\usepackage{bm}
\usepackage{graphicx}
\usepackage{amsmath,amssymb}

\begin{document}
 
  \title{Universal Bound States in Long-range Spin Chains with an Impurity }

  \author{Ning Sun}
  \affiliation{Department of Physics, Fudan University, Shanghai, 200438, China}

  \author{Lei Feng}
  \thanks{leifeng@fudan.edu.cn}
  \affiliation{Department of Physics, Fudan University, Shanghai, 200438, China}
  \affiliation{State Key Laboratory of Surface Physics, Fudan University, Shanghai, 200438, China}
  \affiliation{Institute for Nanoelectronic devices and Quantum computing, Fudan University, Shanghai, 200438, China}
  \affiliation{Shanghai Key Laboratory of Metasurfaces for Light Manipulation, Shanghai, 200433, China}
  \affiliation{Hefei National Laboratory, Hefei 230088, China}

  \author{Pengfei Zhang}
  \thanks{PengfeiZhang.physics@gmail.com}
  \affiliation{Department of Physics, Fudan University, Shanghai, 200438, China}
  \affiliation{State Key Laboratory of Surface Physics, Fudan University, Shanghai, 200438, China}
  \affiliation{Hefei National Laboratory, Hefei 230088, China}

  \date{\today}

  \begin{abstract}
  Understanding how quasi-particles interact with impurities is crucial for unveiling novel properties of quantum many-body systems. A prominent example is the enhanced scattering between electrons and magnetic impurities in the low-energy limit, which gives rise to the Kondo effect. In this letter, motivated by recent developments in quantum simulation platforms, we investigate the universal behavior of long-range quantum spin chains with a single local impurity, focusing on systems that conserve the magnon number. Using effective field theory, we show that distinct classes of universal three-magnon states can emerge when the impurity-mediated two-magnon interaction is on resonance. When the long-range coupling decays as $1/r^\alpha$, we find that (i) for $\alpha\in (2,2.89)$, the system exhibits Efimov effects, with the three-body binding energy forming a geometric series $\ln|E^{(n)}_{\text{3-body}}|\sim-n (\alpha-1) \pi/s_0(\alpha)$, (ii) for $\alpha=2$, the system shows semi-super Efimov effects with $\ln|E^{(n)}_{\text{3-body}}|\sim-(n\pi-\theta)^2/8$. Our theoretical prediction is validated by the numerical solution of the Skorniakov-Ter-Martirosian equation. Our results could be tested experimentally in the future on quantum simulation platforms.
  \end{abstract}
    
  \maketitle

  \emph{ \color{blue}Introduction.--} The interaction between quasi-particles and impurities plays a crucial role in the emergence of novel quantum phases, which in turn governs the macroscopic properties of quantum many-body systems, including charge and thermal transport. A celebrated example is the Kondo effect \cite{Kondo:1964nea,RevModPhys.47.773,hewson1997kondo,Affleck:1995ge}, where a magnetic impurity interacts with the conduction electrons in a metal. It gives rise to the formation of a many-body bound state known as the Kondo singlet. This singlet state, where the impurity spin couples with the surrounding electron spins, results in a characteristic low-temperature behavior, marked by a minimum in resistance. Recent developments have also investigated the impurity problem from the perspective of information scrambling \cite{PhysRevB.110.035137,PhysRevLett.133.266503,PhysRevB.110.235110}, unveiling a universal dynamical transition from a scrambling phase, where the operator size grows persistently, to a non-scrambling phase, where the operator quickly escapes from the impurity.

  \begin{figure}[t]
    \centering
    \includegraphics[width=0.95\linewidth]{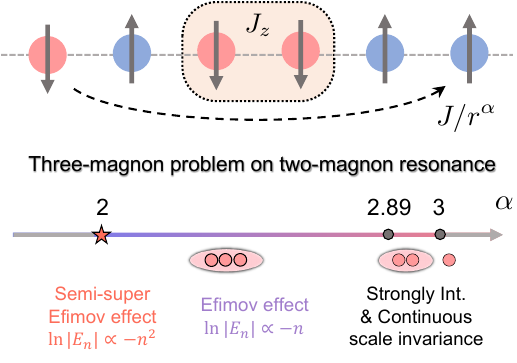}
    \caption{We present a schematic of a long-range spin chain with an impurity. The Hamiltonian includes long-range couplings of the form $J/r^\alpha$ in the $x$- and $y$-directions, as well as a localized interaction of strength $J_z$ along the $z$-direction near the origin. We adopt $\ket{\uparrow}$ as the vacuum state and $\ket{\downarrow}$ represents magnons. By tunning $J_z$ to the two-magnon resonance, the system can exhibit the semi-super Efimov effect for $\alpha=2$ or Efimov effect for $\alpha\in(2,2.89)$. }
    \label{fig:schemticas}
  \end{figure}

  In these studies, the main focus has been on scenarios where the dynamical exponent $z=1$ or $z=2$, which are realized in fermionic systems near the Fermi surface or at the band edge. On the other hand, recent developments in quantum simulation platforms provide a natural realization of synthetic quantum systems with arbitrary dynamical exponents $z$ due to long-range couplings \cite{LEPORI201635,PhysRevB.94.125121,PhysRevA.93.043605,Toappear}. For example, state-of-the-art trapped ion systems \cite{wu2023research,2019ApPRv...6b1314B,Foss-Feig:2024blk,2021NatRM...6..892B,10164086,RevModPhys.93.025001,RevModPhys.93.025001} exhibit long-range couplings with a tunable decay power $\alpha \in (0,3)$ \cite{RevModPhys.93.025001}, covering $z\in(0,2)$ for magnon excitations. In \cite{Toappear}, it is shown that this new ingredient enables the emergence of universal Efimov bound states \cite{EFIMOV1970563,Hammer:2005bp,Nielsen:2001hbm,Ferlaino2010FortyYO,Nishida:2012hf,Braaten:2006vd,Hammer:2010kp,Naidon:2016dpf,Kievsky:2021ghz} in arbitrary dimensions. Specifically, this requires $\alpha \in (1.52,1.88)$ in one-dimensional systems, which aligns well with the parameters of trapped ion systems. Nevertheless, the universal properties of long-range spin chains after introducing impurities remain unexplored.
  
  In this letter, we demonstrate the enrichment of universal few-magnon states in long-range spin chains when an impurity is introduced, represented as an additional two-magnon interaction $J_z$ near the origin, as illustrated in Fig. \ref{fig:schemticas}. In the two-magnon sector, the system exhibits emergent scale invariance in the low-energy limit when tuned to resonance for $\alpha \in (2,3)$. This scale invariance is further broken into discrete scale invariance in the three-magnon sector for $\alpha \in (2,2.89)$, where an infinite tower of Efimov bound states emerges, with the binding energy forming a geometric series. More interestingly, at $\alpha = 2$, the scale factor diverges, and the Efimov effect is replaced by the semi-super Efimov effect \cite{PhysRevLett.118.230601,PhysRevA.96.030702}, with $\ln |E^{(n)}_{\text{3-body}}|\propto -n^2$, which has not been reported in one-dimensional systems. Our results highlight long-range spin models as a novel platform for studying impurity problems with a general dynamical exponent and are of high experimental relevance.

  \emph{ \color{blue}The model.--} To be concrete, we consider a spin-1/2 quantum spin chain with long-range XY couplings. An impurity is added near $n=0$, which is modeled as a local interaction in the $z$-direction. The Hamiltonian reads
  \begin{equation}\label{eqn:microscopic}
  H=-\sum_{n<m}\frac{J}{|m-n|^\alpha}\left(\sigma_n^x\sigma_{m}^x+\sigma_n^y\sigma_{m}^y\right)-J_{z}P_0^\downarrow P_1^\downarrow.
  \end{equation}
  Here, $\sigma_n^l$ with $l \in \{x, y, z\}$ represents the Pauli matrix on the $n$-th site and $P_n^\downarrow=(1-\sigma^z_n)/2$. We introduce the quasi-particle description of the system by taking $\ket{\uparrow}$ as the vacuum state and $\ket{\downarrow}$ as a magnon excitation. In this picture, the spin-rotation symmetry along the $z$-direction corresponds to magnon number conservation, which enables the study of few-magnon states. 

  Since the impurity term contributes only when both sites $0$ and $1$ are occupied by magnons, the single-magnon sector remains translation-invariant. The eigenstate with momentum $k$ is given by $|k\rangle \sim \sum_n e^{ikn} \sigma^-_n |0\rangle$, with excitation energy matching the result obtained without the impurity \cite{Toappear}
  \begin{equation}\label{eq:dispersion}
  \epsilon_k= -\sum_{r=1}^\infty\frac{4J}{r^\alpha} \cos(kr)\approx \epsilon_0+u |k|^z.
  \end{equation}
  Here, we assume $\alpha>1$ and only keep the leading-order contribution at small $k$. The dynamical exponent is $z = \min\{2, \alpha - 1\}$, and the detailed expression for $u$ will not be important. For conciseness, we will drop the constant part $\epsilon_0$, which can always be set to zero by adding a homogeneous magnetic field. 

  When considering states with more than one magnon, there are two different types of interactions. Firstly, there cannot be two magnons on a single site, which is equivalent to an infinite on-site repulsion. However, this repulsive interaction cannot induce a two-body resonance \cite{landau2013quantum,zhai2021ultracold}, where a bound state would emerge. Secondly, for $J_z > 0$, the impurity term induces an attractive interaction between magnons near the origin. A two-magnon resonance can be induced by the impurity when tuning $J_z/J$, which plays a dominant role in low-energy physics. With these understandings, we introduce the effective field theory that captures the universal low-energy scattering between magnons
  \begin{equation}\label{eq:fieldtheory}
  L=\sum_k \bar{\psi}_k(i\partial_t-\epsilon_k )\psi_k-\frac{g}{2}\bar{\psi}(0)\bar{\psi}(0){\psi}(0){\psi}(0).
  \end{equation}
  Here, the interaction term only presents at $x = 0$. The theory is defined with a large-momentum cutoff $\Lambda$. We tune $g$ to achieve a two-magnon resonance, as elaborated in discussions below. For later convenience, we perform the Hubbard-Stratonovich transformation to introduce the dimer field $d = g \psi(0)\psi(0)$. This replaces the four-magnon interaction with 
  \begin{equation}\label{eqn:dimer}
  -\frac{1}{2} \left(\bar{\psi}(0)\bar{\psi}(0) d + \bar{d} \psi(0)\psi(0) - \frac{\bar{d}d}{g} \right).
  \end{equation}
  Notice that although we focus on the microscopic Hamiltonian \eqref{eqn:microscopic}, the effective field theory is generally valid for a large class of quantum systems with magnon number conservation and local impurities, making predictions that are universal.

  \emph{ \color{blue}Efimov Effect.--} We first consider scenarios where the system can exhibit the Efimov effect, beginning with the analysis of the two-magnon problem. Due to the lack of translation symmetry, the only conserved quantity is the total energy $E$. In the dimer field picture, the scattering $T$-matrix between two incoming magnons is proportional to the propagator of the dimer field, which is given by
  \begin{equation}\label{eqn:Tmatrix}
  T(E)^{-1}=\frac{1}{2g}-\frac{1}{2}\int_{-\Lambda}^\Lambda\frac{dq_1dq_2}{(2\pi)^2}\frac{1}{E_+-\epsilon_{q_1}-\epsilon_{q_2}}.
  \end{equation}
  Here, we have introduced the notation $E_+=E+i0^+$. The second term represents the contribution from a virtual process where a dimer field splits into two magnons, with momenta $q_1$ and $q_2$. Compared to standard scattering theory in one dimension, the momenta of the two magnons become independent and are integrated individually. Efimov bound states can emerge in the three-magnon problem only if the resonant two-magnon scattering satisfies (i) scale invariance and (ii) strong interactions, as fulfilled in three-dimensional non-relativistic particle systems \cite{Hammer:2005bp}. Mathematically, this requires the presence of a single divergent term in the integral as $\Lambda \to \infty$ \cite{Toappear}, which requires $z\in (1,2)$. Under this constraint, we find 
  \begin{equation}\label{eqn:two-body}
  T(E)^{-1}=G^{-1}-B(z)u^{\frac{2}{z}}(-E_+)^{\frac{2}{z}-1},
  \end{equation}
  where 
  \begin{equation}
      B(z)=-\frac{2^{-\frac{z+2}{z}} \csc \left(\frac{\pi }{z}\right) \Gamma
   \left(\frac{1}{2}-\frac{1}{z}\right) \Gamma \left(\frac{1}{z}\right)}{\pi ^{3/2} z^2}
  \end{equation}
   and we have introduced the low-energy scattering parameter $G$ through the renormalization relation 
   \begin{equation}
   \frac{1}{G}\equiv \frac{1}{2g}+\frac{ \csc \left(\frac{\pi }{z}\right)}{2 \pi  u z(2-z)}\Lambda ^{2-z}.
   \end{equation}
   The two-magnon scattering becomes resonant at $G = \infty$, where a bound state emerges at the two-magnon threshold. In this case, the two-magnon sector exhibits continuous scale invariance, indicating the possibility of an Efimov effect in the three-magnon sector. More generally, a shallow bound state exists when $G>0$, with a binding energy $|E_{\text{2-body}}|=(GB(z)u^{\frac{2}{z}})^{-(2-z)/z}$.

  \begin{figure}[t]
    \centering
    \includegraphics[width=0.85\linewidth]{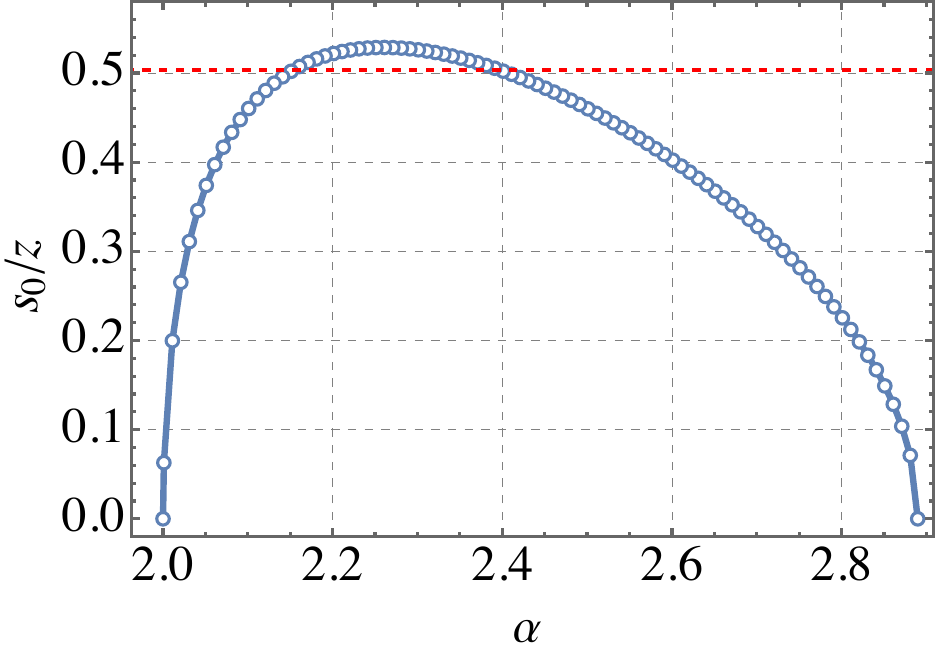}
    \caption{We present the numerical solution of Eq. \eqref{eqn:resEfimov} as a function of $\alpha=1+z$, which gives the scale factor for Efimov states $E^{(n)}_{\text{3-body}}=E^{(0)} \exp({-\frac{(\alpha-1)\pi}{s_0}}n)$. The results show the existence of Efimov states for $\alpha \in (2,2.89)$. The red dashed line represents the known result for identical bosons in three dimensions. }
    \label{fig:numEfimov}
  \end{figure}

   We proceed to study the three-magnon problem at $G = \infty$ using the Skorniakov-Ter-Martirosian (STM) equation \cite{Skorniakov:1957kgi}. Focusing on the three-body bound state with total energy $E < 0$, the bound-state magnon-dimer wavefunction $\varphi(E,k)$ satisfies
   \begin{equation} \label{eqn:STM}
   \varphi(E,k)=\int \frac{dq}{2\pi}\int \frac{dq_2}{2\pi} \frac{T(E-\epsilon_q)}{E-\epsilon_k-\epsilon_q-\epsilon_{q_2}}\varphi(E,q).
   \end{equation}
   The exhibition of Efimov effect can be detected by solving the wavefunction at $E=0$ \cite{Hammer:2005bp}. Denoting $\varphi(k)\equiv \varphi(0,k)$, the integral equation takes the form of
   \begin{equation}\label{eqn:E=0}
   \varphi(k)=\frac{\csc(\frac{\pi}{z})}{z B(z)}\int_{-\infty}^{\infty} \frac{dq}{2\pi} \frac{1}{|q|^{2-z}}\left[\frac{1}{|k|^z+|q|^z}\right]^{1-\frac{1}{z}} \varphi(q),
   \end{equation}
   where we have carried out the integration over $q_2$. Eq. \eqref{eqn:E=0} is invariant under the scale transformation $(k, q) \rightarrow \lambda (k, q)$. As a consequence, the equation is solved by the ansatz $\varphi(k) = |k|^{s_1 + i s_0}$. Substituting this ansatz into Eq. \eqref{eqn:E=0}, we obtain $s_1=\frac{1-z}{2}$, and $s_0$ is determined by 
   \begin{equation}\label{eqn:resEfimov}
   {\left|\Gamma \left(\frac{-2 i s_0+z-1}{2 z}\right) \right|^2}= \Gamma \left(1+\frac{1}{z}\right) \Gamma \left(-\frac{2}{z}\right).
   \end{equation}
   Here, $\Gamma(n) \equiv \int_0^\infty x^{n-1} e^{-x} dx$ is the Gamma function. The numerical solution of Eq. \eqref{eqn:resEfimov} is presented in Fig. \ref{fig:numEfimov}, which exists for $z \in (1, 1.89)$. This corresponds to $\alpha \in (2, 2.89)$. In this regime, the zero-energy wavefunction is given by $\varphi(k)\sim |k|^{s_1}\cos(s_0\ln k/\Lambda_*)$, where $\Lambda_*$ is an additional three-body parameter \cite{Bedaque:1998kg, Bedaque:1998km}. This indicates that continuous scale invariance is broken to discrete scale invariance, since a scale transformation $k \rightarrow \lambda k$ leaves the form of the wavefunction unchanged only when $\lambda = e^{-\frac{n\pi}{s_0}}$. Consequently, the binding energy of all three-body bound states should be organzied into a geometric series, with a scale factor $e^{-z{\pi}/{s_0}}$. This is directly demonstrated by numerically solving the STM equation \eqref{eqn:STM} with the $T$-matrix in Eq. \eqref{eqn:two-body} on two-body resonance. The results are presented in TABLE \ref{TAB:numeric}, which matches our theoretical prediction to good accuracy. We also confirm the absence of shallow three-body bound states for $\alpha \in (2.89, 3)$, where the three-body problem maintains the continuous scale invariance, as illustrated in FIG. \ref{fig:schemticas}. 

  \begin{table}[t]
  \centering 
  \begin{tabular}{|c|c|c|c|c|c|}
    \hline
    \hline
    &\multicolumn{2}{c|}{$\alpha$=2.2} &\multicolumn{2}{c|}{$\alpha$=2.5}\\ 
    \hline
    $n$ & $\phi_n$  & $\phi_n-\phi_{n-1}$& $\phi_n$  & $\phi_n-\phi_{n-1}$ \\ 
    \hline
    1  & 1.837 &   & 2.238 &    \\ 
    2 & 3.750 &1.913 &  4.405 &2.167\\ 
    3  & 5.663 &1.913 & 6.572 &2.167  \\ 
    4  & 7.576&1.913 &  8.740 &2.168  \\ 
   \hline
    $z/s_0$ & & 1.917& & 2.173 \\
   \hline   
    \end{tabular}
  
  \caption{We present the three-magnon binding energies, parametrized as $E^{(n)}_{\text{3-body}}=u\Lambda^{z}\exp(-\pi\phi_n)$, obtained by numerically solving Eq. \eqref{eqn:STM} with the $T$-matrix \eqref{eqn:two-body} on the two-body resonance $P\rightarrow \infty$. The results clearly demonstrate the presence of Efimov bound states, with $\phi_{n}-\phi_{n-1}$ matching the theoretical prediction shown in Fig. \ref{fig:numEfimov} with good accuracy.  }\label{TAB:numeric}
\end{table}

  \emph{ \color{blue}Semi-super Efimov Effect.--} The discussion above assumes the validity of Eq. \eqref{eqn:two-body}, and is therefore limited to $\alpha \in(2,3)$. Interestingly, the Efimov effect persists as we tune $\alpha\rightarrow 2$, raising a new question about the fate of the Efimov bound states when $\alpha=2$. In this section, we show that there is still an infinite tower of bound states, although organized in a different pattern, referred to as the semi-super Efimov effect \cite{PhysRevLett.118.230601,PhysRevA.96.030702}. Our results provide the first realization of the semi-super Efimov effect in quantum spin chains.

  We begin by revisiting the two-body calculation for $\alpha=2$. The integral in Eq. \eqref{eqn:Tmatrix} now contains more than one divergent term, which can not be renormalized by a single bare parameter $g$. This situation is similar to $p$-wave scattering in three dimensions \cite{PhysRevLett.94.230403,PhysRevA.86.012711}, where a two-channel model is required for renormalization. In a similar spirit, we construct a two-channel model with a dynamical dimer field as
  \begin{equation}\label{eqn:dimer2}
  \begin{aligned}
  L=&\sum_k \bar{\psi}_k(i\partial_t-\epsilon_k )\psi_k+\bar{d}(i\partial_t-\nu_0)d\\
  &-{g_0}\left(\bar{\psi}(0)\bar{\psi}(0) d + \bar{d} \psi(0)\psi(0) \right)/2.
  \end{aligned}
  \end{equation}
  The $T$-matrix of two incoming mganons with total energy $E$ contains additional contribution from the kinetic term of dimer fields, which gives
  \begin{equation}\label{eqn:Tmatrix2}
  T(E)^{-1}=\frac{E}{g_0^2}-\frac{\nu_0}{g_0^2}-\frac{1}{2}\int_{-\Lambda}^\Lambda \frac{dq_1dq_2}{(2\pi)^2}\frac{1}{E_+-\epsilon_{q_1}-\epsilon_{q_2}}.
  \end{equation}
  As a consequence, the renormalized $T$-matrix reads
  \begin{equation}\label{eqn:Tsemi}
  T(E)^{-1}=\frac{1}{a}-\frac{1}{2\pi^2 u^2} E_+\ln (-E_+ R/u).
  \end{equation}
  Here, we introduce the scattering length $a$ and effective range $R$, which are related to $\nu_0$ and $g_0$ through the renormalization relation
  \begin{equation}
  \begin{aligned}
  &\frac{1}{a}=-\frac{\nu_0}{g_0^2}+\frac{\ln 2 }{\pi^2 u}\Lambda,\\
  &\frac{2\pi^2 u^2}{g_0^2}=-\ln \left(\frac{R\Lambda e}{2}\right).
  \end{aligned}
  \end{equation}
  Now, the two-magnon system exhibits a shallow bound state when $a>0$, with binding energy $|E_{\text{2-body}}|$ determined through the pole of the $T$-matrix. A two-magnon resonance corresponds to tunning $a\rightarrow \infty$, while leaving an order one effective range $R\sim O(1)$ \footnote{This is because we have chosen the lattice constant $a = 1$. Otherwise, we expect $R \sim a$ near the two-magnon resonance, which corresponds to the range of the microscopic potential.}. Consequently, the resonant point is actually weakly interacting, characterized by $T(E) \to 0$ as $E \to 0$. In addition, there is no scale invariance due to the presence of the effective range, which prevents the exhibition of the traditional Efimov effect, while suggesting the possible emergence of the super-Efimov effect \cite{PhysRevLett.110.235301,PhysRevA.90.063631,2014arXiv1405.1787G,PhysRevA.92.020504,PhysRevA.95.033611} or the semi-super Efimov effect \cite{PhysRevLett.118.230601,PhysRevA.96.030702}. 

\begin{table}[t]
  \centering 
  \begin{tabular}{|c|c|c|}
    \hline
    \hline
    &\multicolumn{2}{c|}{$\alpha$=2} \\ 
    \hline
    $n$ & $\phi_n$  & $\phi_n-\phi_{n-1}$ \\ 
    \hline
    1  & 1.166 &   \\ 
    2 & 2.638 &1.472 \\ 
    3  & 3.796 &1.158  \\ 
    4  & 4.883 &1.087 \\ 
    5  & 5.938 &1.055 \\ 
    6  & 6.977 &1.039 \\
   \hline
    Theory  & & 1\\
   \hline   
    \end{tabular}
  
  \caption{We present the three-magnon binding energies, parametrized as $E^{(n)}_{\text{3-body}}=u\Lambda \exp(-\pi^2\phi_n^2/8)$, obtained by numerically solving Eq. \eqref{eqn:STM} with the $T$-matrix \eqref{eqn:Tsemi} on the two-body resonance $a\rightarrow \infty$ and $\Lambda R=0.5$. The results demonstrate the presence of semi-super Efimov bound states, with $\phi_{n}-\phi_{n-1}$ approaching the theoretical prediction with good accuracy.  }\label{TAB:numeric2}
\end{table}

  Three-body bound states are still determined by solving the STM equation \eqref{eqn:STM} with the modified two-body $T$-matrix \eqref{eqn:Tsemi}. We again focus on the zero-energy wavefunction $\varphi(k)$, which now satisfies ($k>0$)
  \begin{equation}
  \varphi(k)=-\int_0^\Lambda dq\frac{2}{q \ln (q R)}\ln\left(\frac{\Lambda}{q+k}\right)\varphi(q).
  \end{equation}
  This equation is solved analytically using the leading-logarithm approximation \cite{PhysRevD.59.094019,PhysRevLett.110.235301,PhysRevLett.118.230601}. We split the integral into two parts $0<q<k$ and $k<q<\Lambda$, and approximate $q+k$ as $\text{max}\{q,k\}$. This transforms the integral equation into a differential equation
  \begin{equation}
   \frac{d}{dk}\left[k\frac{d}{dk}\varphi(k)\right]=\frac{2}{k\ln(kR)}\varphi(k).
  \end{equation}
  Introducing $w=-\ln(kR)$, the solution is given by $\varphi(k)\sim \sqrt{w}[c_1J_1(2\sqrt{2w})+c_2Y_1(2\sqrt{2w})]$, where $J_n(x)$ and $Y_n(x)$ are Bessel functions of the first and second kinds, respectively. In the low-energy regime, $k\ll 1/R$, we obtain the asymptotic behavior $\varphi(k)\sim w^{1/4}\sin(2\sqrt{2w}+\theta)$, with a three-body parameter $\theta$. Similar to the Efimov wavefunction, the result exhibits $n\pi$ periodicity in $2\sqrt{-2\ln(kR)}$, implying an infinite number of three-magnon bound states with binding energies $\ln|E^{(n)}_{\text{3-body}}|\sim-(n\pi-\theta)^2/8$ in the limit of $n\rightarrow \infty$. This is known as the semi-super Efimov effect, originally discovered for two-dimensional bosons with resonant three-body interactions \cite{PhysRevLett.118.230601}. To validate the theoretical prediction, we numerically determine the binding energies by solving the STM equation \eqref{eqn:STM} with the $T$-matrix in \eqref{eqn:Tsemi} at $a=\infty$. The results for $\Lambda R=0.5$ are presented in TABLE \ref{TAB:numeric2}.

  \emph{ \color{blue}Discussions.--} In this letter, we study the universal low-energy scattering problem of a few magnons in long-range spin chains. By applying effective field theory, we reveal the universal features of the three-magnon problem when the two-magnon interaction is on resonance. Firstly, for $\alpha \in (2,2.89)$, there exists an infinite tower of Efimov bound states with discrete scale invariance, where the binding energies are organized in a geometric series $\ln|E^{(n)}_{\text{3-body}}|\sim-n (\alpha-1) \pi/s_0(\alpha)$. Secondly, for $\alpha=2$, the system exhibits instead the semi-super Efimov effect, with binding energies $\ln|E^{(n)}_{\text{3-body}}|\sim-(n\pi-\theta)^2/8$ in the limit of $n\rightarrow \infty$. Finally, for $\alpha \in (2.89,3)$, there is no three-magnon bound state, indicating that the system maintains its continuous scale invariance. We expect these results to be validated experimentally in the future on quantum simulation platforms.

  While we focus on one-dimensional spin chains, it is interesting to generalize our analysis to higher-dimensional systems, where the effect of defects with arbitrary codimensions can be studied. This is analogous to previous studies of non-relativistic particles in mixed dimensions \cite{Nishida:2011ew,PhysRevLett.101.170401,PhysRevA.79.060701}. Since both the Efimov effect and the semi-super Efimov effect have been identified in our work, it remains unknown whether the super-Efimov effect \cite{PhysRevLett.110.235301,PhysRevA.90.063631,2014arXiv1405.1787G,PhysRevA.92.020504,PhysRevA.95.033611}, or other classes of universal bound states, manifest in quantum spin systems. This may require the study of the three-magnon problem near a high partial-wave resonance. Finally, universal few-body physics has broad implications for many-magnon states, particularly when the magnon density is low but finite. In particular, a series of contact relations can be derived, as in ultra-cold atomic gases \cite{TAN20082971,PhysRevLett.100.205301,PhysRevLett.99.190407,PhysRevLett.99.170404,PhysRevLett.116.045301,2009EPJB...68..401W,PhysRevA.79.053640,PhysRevA.79.023601}. We leave these studies for future work.

\textit{Acknowledgement.} 
This project is supported by the Shanghai Rising-Star Program under grant number 24QA2700300 (PZ), the NSFC under grant 12374477 (PZ), the Innovation Program for Quantum Science and Technology 2024ZD0300101 (PZ) and 2023ZD0300900 (LF), and the Shanghai Municipal Science and Technology Major Project grant 24DP2600100 (NS and LF).

\bibliography{Impurity.bbl}

\end{document}